\documentclass[epjST]{svjour}
\usepackage{graphics}
\usepackage[utf8]{inputenc}
\begin{document}
\title{On salesmen and tourists: two-step optimization in deterministic foragers}
\author{ M Maya\inst{1}  \& O Miramontes\inst{1,2}  \& D Boyer\inst{1}}
%
\institute{$^1$Instituto de F\'{i}sica and C3, Universidad Nacional Aut\'{o}noma 
de M\'{e}xico, Apartado Postal 20-364, 01000 Mexico CDMX, M\'{e}xico \& 
$^2$Departamento de Matem\'{a}tica Aplicada a la Ingenier\'{i}a Aeroespacial, 
ETSI Aeron\'{a}uticos, Universidad Polit\'{e}cnica de Madrid, Madrid,
Spain.}
%


\abstract{We explore a two-step optimization problem in random environments, the so-called restaurant-coffee shop problem, where a walker aims at visiting the nearest and better restaurant in an area and then move to the nearest and better coffee-shop. This is an extension of the Tourist Problem, a one-step optimization dynamics that can be viewed as a deterministic walk in a random medium. A certain amount of heterogeneity in the values of the resources to be visited causes the emergence of power-laws distributions for the steps performed by the walker, similarly to a L\'{e}vy flight. The fluctuations of the step lengths tend to decrease as a consequence of multiple-step planning, thus reducing the foraging uncertainty. We find that the first and second steps of each planned movement play very different roles in heterogeneous environments. The two-step process improves only slightly the foraging efficiency compared to the one-step optimization, at a much higher computational cost. We discuss the implications of these findings for animal and human mobility, in particular in relation to the computational effort that informed agents should deploy to solve search problems.  
} 
\maketitle
\section{Introduction}
\label{intro}

The Traveling Salesman Problem (TSP) is paradigmatic of combinatorial optimization problems and has far reaching applications in many fields of science. In the TSP, an agent must find the shortest route that visits a set of $N$ cities, each city being visited once, before returning to its starting point. This is a NP-hard problem due to the global nature of the optimization, and the computation time to find the solution is of the order of $N!$. On the other hand, the Tourist Problem (TP) is a {\it local} optimization problem where a visitor goes from one city to the nearest-neighboring city, and so on \cite{Stanley2,Lima}. Global and local optimization problems are of great importance in physics since these are related to variational problems, with applications to the determination of Hamiltonian parameters in spin systems, for instance \cite{Lunga}. 

The TP can be considered as a deterministic walk that explores a disordered medium (if the cities are distributed randomly, for instance) by taking steps as short as possible given a discrete number of possibilities at each step \cite{Lima}. Therefore, the walker's driving force is a local optimization procedure, rather than a stochastic force as in standard random walks. In this case, the walker has some information about the environment and uses that information to take presumably efficient movement decisions. As a matter of fact, many animal species \cite{fagan2013spatial} and humans \cite{song2010limits} keep memory of their previous activities and can be considered as deterministic to some extend. These cognitive abilities can be used, for instance, to maximize a foraging efficiency, {\it e.g.}, the amount of visited resources per unit of traveled distance, a problem that the TSP and the TP tackle with different levels of complexity.


Many animals, when searching for resources, are known to perform trajectories that are reminiscent of L\'evy flights  \cite{Viswanathan1}, see also \cite{reynolds2015liberating} and references therein. In empirical data, the statistics of displacement lengths are often approximated by power-laws, which are distributions that lack a characteristic scale and are dominated by rare, very long jumps. The common interpretation for the presence of these patterns in ecology is that L\'evy walk movements optimize the success of random searches when foragers have no information about resources, and when those are scarce and randomly distributed in patches \cite{Viswanathan2,Viswanathan3}. A less studied interpretation of biological L\'evy flights is that the forager responds to a complex distribution of resources, which induces movements that reflect the environmental heterogeneity. According to this hypothesis, L\'evy walks are no longer an internal search process but emerge from an ecological interaction \cite{Miramontes3,Miramontes4,Boyer3}. Along this line, L\'evy patterns can emerge if animals follow mental maps that contain information about the location and quality of heterogeneous food resources \cite{Viswanathan2,Miramontes1,Brockmann,Bartumeus}. Evidence actually supports the use of memory in monkeys  \cite{Miramontes1,Boyer2,Boyer3}, humans \cite{Barabasi,Baronchelli} and many other animal species. 

In many foraging theories that incorporate information use (see also {\it e.g.} \cite{fronhofer2013random,bracis2015memory}), each movement decision usually follows, like in the TP, a single-step rule. Typically, the forager evaluates the best move, given its current position and environmental conditions. In this paper, we wish to extend this reasoning to the less-studied multiple-step optimization processes, where a forager evaluates at once the outcome of a combination of several steps. This is an intermediate case between the TP and the TSP, but still of much lower computational difficulty than the latter. Considering here the simplest case of two-step planning, we study how the foraging efficiency of the walker is improved, and how the L\'evy patterns emerging in the single-step processes are affected by the introduction of the second step.


\section{Model}\label{model}

In this  paper we  extend the model introduced originally in  \cite{Boyer3}, that was motivated by observations on fruit-eating monkeys. We consider a two  dimensional squared  domain of unit area filled with $N$ randomly and uniformly distributed  point-like targets at fixed positions.  Each  target $i$  has  a size or attractiveness  $k_i$,  which is a random variable drawn from a given distribution $p(k)$. In this disordered environment, we consider a forager with a perfect knowledge of the sizes and positions of the targets. Ref. \cite{Boyer3} studied trajectories generated by a one-step optimization rule, that we call from now on the model ‘$o1p$’. In this rule, the forager located, say, on target $i$ chooses the next target to visit, $j^*\ne i$, such that the distance between $i$ and $j^*$ divided by the size of $j^*$ is minimal. Therefore $j^*$ minimizes the cost function

\begin{equation}
\label{cost1}
E^{(1)}=l_{ij} / k_j,
\end{equation}

\noindent where $l_{ij}$ is the distance between targets $i$ and $j$, and where $j$ cannot be an already visited target. The above process is iterated, generating a deterministic trajectory which does not revisit twice a same target. With Eq. (\ref{cost1}), the forager aims at obtaining as many resources as possible (large $k_i$'s) in the shortest traveled distance.

The model studied here is a two-step version of the one above, and is referred to as ‘$o2p$’ in the following. To take an image, this model describes the situation of a person that wants to visit a good restaurant not too far away, followed from there by a trip to a nearby good coffee shop. Or, to go to a cheap gas station, from where a nearby cash machine can be visited afterwards, etc…    

\begin{figure}
\centering
\includegraphics{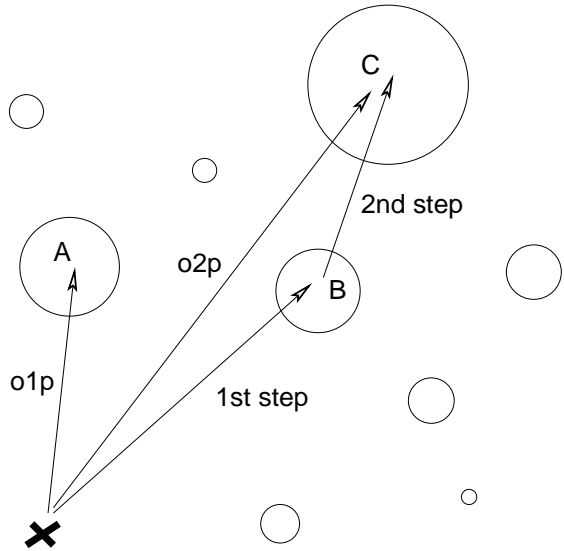}
\caption{First choice of a forager that performs a one-step ($o1p$) and two-step 
         optimization ($o2p$). The forager starts at the location indicated by the cross.}
\label{fig:modelo}
\end{figure}

This model assumes, like in \cite{Boyer3}, that the resource sizes are distributed according to a power-law distribution:

\begin{equation}
\label{disttam}
p(k) = C k^{-\beta}, \qquad \mbox{con} \qquad k = 1,2,\dots, k_{\mathrm{max}}, 
\end{equation}

\noindent with $k$ an integer, $k_{\mathrm{max}}$ a cutoff, $C$  is the  normalization  constant,  and  $1<\beta <\infty$  is  the parameter which  characterizes the heterogeneity of the medium regarding the size of the targets.  The assumption of  the power-law  form  is based  on evidence that resources, in particular fruit trees on which many animals feed, are often distributed along fat-tailed distributions \cite{Enquist,Newmann2005}.  The forager is initially located at the middle of the domain and then moves from one target to  another. At  each  step, the  following rules  are iterated:

$(i)$  The  forager located, say, at  target  $i$  considers an available  target (not visited before),  $j \neq  i  $;

$(ii)$ then, the forager considers a second available target, $m\ (\neq i,
j)$;

$(iii)$ among all possible pairs ($j,m$) of available targets, the  forager will visit the one ($j^*$, and then $m^*$), such that the following cost function is minimal:

\begin{equation} 
\label{funec2}
E^{(2)} = \frac{l_{ij} + l_{jm}}{k_j + k_m}.
\end{equation}
    
\noindent Like Eq. (\ref{cost1}), this expression represents, for the combined steps, the distance traveled per unit of resources obtained.
    
$(iv)$ Previously visited targets are not revisited.

The rules of the models $o1p$ and $o2p$ are depicted in Figure \ref{fig:modelo}. We notice that, given a same medium and starting point, the sites chosen may differ in principle in both rules. For instance, the most important target $C$ is closer to target $B$ than to target $A$, therefore, using $o2p$, $B$ is preferred in route to $C$, despite of the fact that $B$ is less important than $A$, which is chosen first in $o1p$.
   
We can distinguish, roughly speaking, three different kinds of environment. $(i)$ When $\beta < 2 $ there are many targets with huge sizes, and these environments belong to the abundant regime. $(ii)$ If  $2 <\beta <  4$, the mean target size does no scale up with $k_{\mathrm{max}}$ but the variance of the size can be large, these environment belong to the so-called diverse regime.  $(iii)$ For $\beta > 4  $ most of  the targets  have the minimal unit size, and we  call this regime the homogeneous or scarce regime.

In the simulations of the model below, we consider $N=10^4$, and the total number of visited targets in one trajectory is $n=500\ll N$. The cutoff value for the maximum target sizes is set to $ k_{\mathrm{max}} = 10^3 $.  Every point depicted in the following figures represents an average over 100 trajectories (each taking place in a different independent medium).


\section{Results} \label{result}

\begin{figure}
\centering
\includegraphics{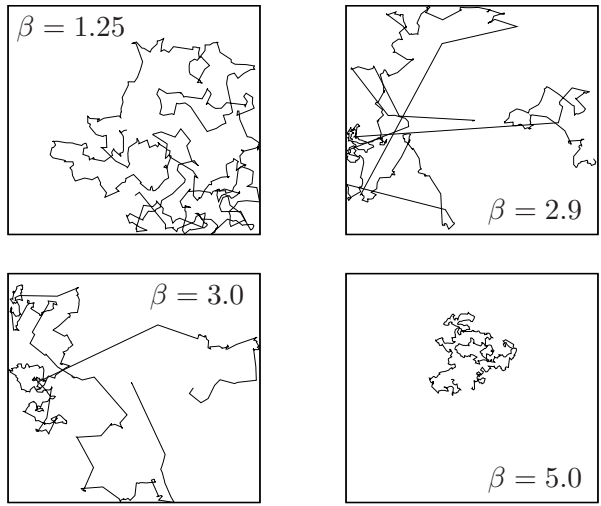}
\caption{Emerging forager trajectories in some media with different values of $\beta$.}
\label{fig:desp}
\end{figure}

Examples of emerging trajectories of a forager in the $o2p$ model are depicted in Figure \ref{fig:desp}. On average, the movement patterns in the domain primarily depend on the resource exponent $\beta$, which is the main parameter in both models. 

It can be noticed that when the  media is abundant,  $\beta < 2 $,  the forager tends to visit large targets. Since those are numerous, there is a characteristic size around which the length of the steps does not fluctuate much (see below). In the case of the  diverse regime,  $2 <  \beta < 4$, the lengths of the steps are much more heterogeneous, which can be interpreted by the fact that relatively few targets have a very large size. Given their relative scarcity, sometimes, it is worth for the forager to perform a very large jump to reach one of them. These movements qualitatively resemble stochastic L\'evy flights, as in the one-step model\cite{Boyer3,Miramontes2}. Finally in the scarce regime,  $\beta > 4$, most of the targets have small sizes, in this case the forager roughly performs a local motion to nearest-neighbor unvisited targets. This case would be the equivalent of a deterministic “Brownian” regime.

\begin{figure}
\centering
\includegraphics{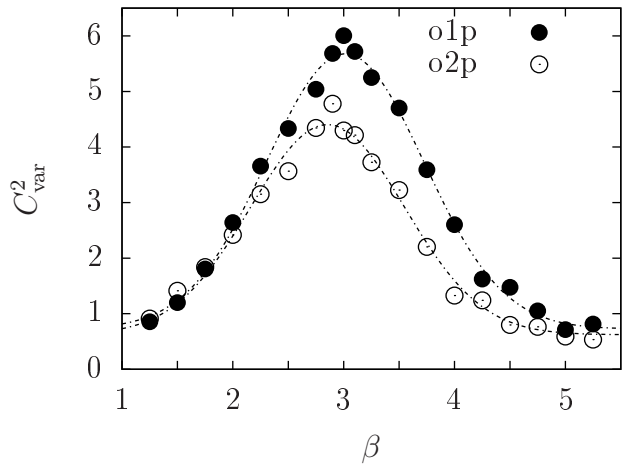}
\caption{Relative fluctuations of the step lengths, as given by $C_{\mathrm{var}}^2$, as a function of the resource exponent $\beta$. Dashed lines are visual guides to the eye only.}
\label{fig:cvar}
\end{figure}

In Figure  \ref{fig:cvar}, we display the  coefficient of variation, that measures the relative fluctuations of the step lengths:

\begin{equation} 
\label{coefvar}
C_{\mathrm{var}} = \frac{\sigma}{\langle l \rangle},
\end{equation}

\noindent where $\sigma$ and $\langle l \rangle$ are the standard deviation and the mean of the  target-to-target jump lengths, respectively. Here, we have aggregated the lengths of the first and the second steps in the two-step process.  The quantity  $  C_{\mathrm{var}}^2$ is useful to compare  the dispersion  between  sets of measures  of different random variables \cite{Black}. In Figure \ref{fig:cvar}, we also show $C_{\mathrm{var}}^2$ in the $o1p$ case for comparison.

There are two results worth noting. First, the two-step optimization model exhibits the largest fluctuations for a particular environmental value, $\beta \approx 3$, which is roughly the same as for the one-step optimization model. In this sense, the media with $\beta \approx 3$ produce the most heterogeneous trajectories. A second remarkable result is that the fluctuations, in a same medium, are weaker in the $o2p$ model. Therefore, by optimizing pairs of steps, the forager reduces the uncertainty on the length of the resulting steps, or, in other words, travels with more similar step lengths than in the single optimization process. One may anticipate that by using three-step, four-step optimizations and so on, the fluctuations are likely to decrease even more. In the limiting case of $N$-step processes, actually, our problem reduces to the standard TSP, since the denominator in Eq. (\ref{funec2}) would be replaced by the total amount of resources, which is independent of the path. In the TSP, the step lengths are known to be short and to fluctuate little.  

\begin{figure}
\centering
\includegraphics{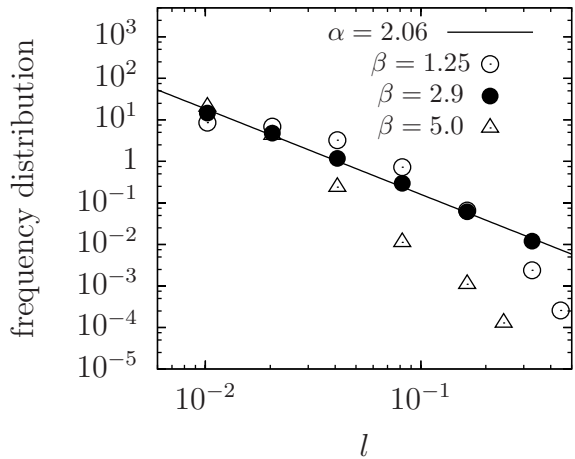}
\caption{Frequency distribution of the jump lengths in the $o2p$ model. A clear power-law behavior is observed for $\beta$ around  $3$.} 
\label{fig:frecdist}
\end{figure}

Here, the relatively large  fluctuations of $l$ near $\beta=3$ are indicative of power-laws distributions. We computed the  frequency distribution $P(l)$ of the target-to-target distances for each step, as depicted in Figure \ref{fig:frecdist} (again grouping the first and second steps of each optimization).  There are  two limits on the lengths in our problem: the  first one is the  characteristic distance  between  nearest-neighbors, of order $1/\sqrt{N}$, and  the second is the  domain size itself (unity). In Figure \ref{fig:frecdist}  we focus on the scales comprised between these two limiting lengths.  Again, like in the $o1p$ case \cite{Boyer3,Miramontes2}, and in sharp contrast with the Poissonian distribution between nearest-neighbor targets in the medium, when $\beta$ is around $3$, the power-law is a good fit for $P(l)$. For $\beta =  2.9$, we determine $P(l) \sim l^{- \alpha}$, with $\alpha = 2.06$. For environments with $\beta$ significantly $>3$ or $<3$, $P(l)$ is poorly described by a power-law.

We now continue with the foraging efficiency of the two-step optimization model, and further compare it with the results of the well-known $o1p$ model \cite{Miramontes2}. The total amount  of resources  captured  after having visited the $n$ targets and the total distance  traveled are determined in each simulation. Obviously, the computational time needed to move from one target to another for a forager doing $o1p$ is much shorter than in the $o2p$ model. 

\begin{figure}
\centering
\includegraphics{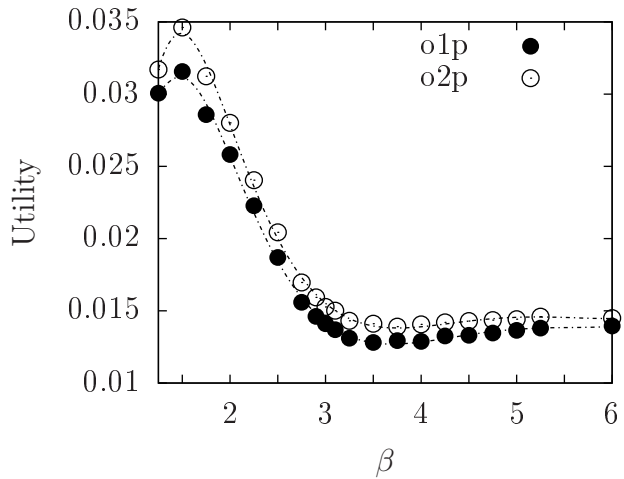}
\caption{Comparison of the utility obtained in the two optimization models.}
\label{fig:util}
\end{figure}

An utility function (synonymous for efficiency) can be defined as 

\begin{equation}
\mathrm{Utility} =\left\langle \frac{K_T}{l_T} \right\rangle,
\end{equation}

\noindent where  $K_T=\sum_{path}k_i/\sum_{i=1,N}k_i$ is the ratio  between the cumulated amount of visited (or "captured") resources and the total amount of resources in the system, $l_T$ is the total distance traveled, and the brackets denote an average over independent realizations. The forager utility is plotted in Figure \ref{fig:util} as a function of $\beta$.

\begin{figure}
\centering
\includegraphics{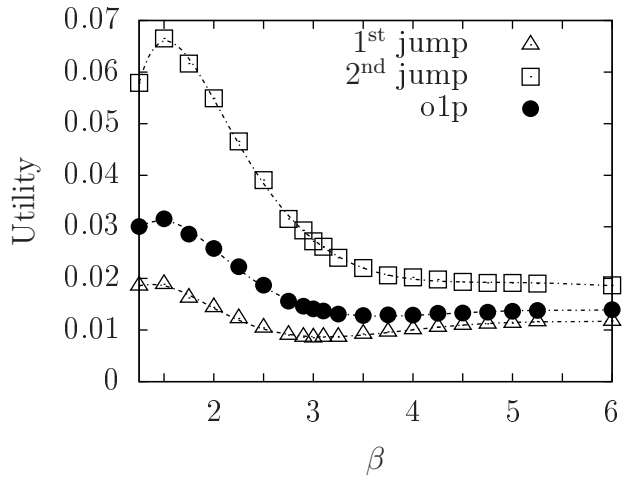}
\caption{Utility obtained by the first and second jumps of the two-step process, compared with that of the single-step model, $o1p$.}
\label{fig:o2putil}
\end{figure}

As one may have expected, the utility achieved by the $o2p$ model is greater than that of the $o1p$ model. However, considering the much larger computing time required in the former (of $O(N^2)$ at each step, instead of $O(N)$), the increase in efficiency is quite marginal. It is instructive, though, to separate the contributions of the two steps during each choice made in the $o2p$ model. Thus, we have calculated separately the average utility of the jumps $i\rightarrow j$ (first jump) and $j\rightarrow m$ (second jump) in the notation above. In figure \ref{fig:o2putil}, the average utility of these jumps is displayed, as well as the  utility of the forager  performing $o1p$ in comparison. There is a remarkable asymmetry between the efficiencies of the  first and the second  jumps in the  $o2p$, the second jump being much more efficient. In addition, the first jump (as illustrated in the cartoon of Figure \ref{fig:modelo}) is {\it less} efficient on average than a $o1p$ jump, which undermines the overall efficiency of the two-step optimization.

\begin{figure}
\centering
\includegraphics{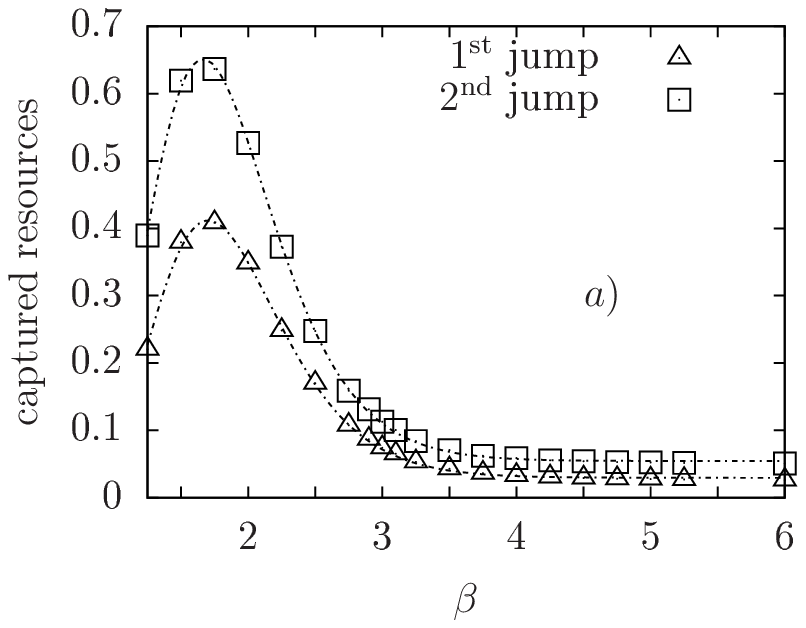} \\
\includegraphics{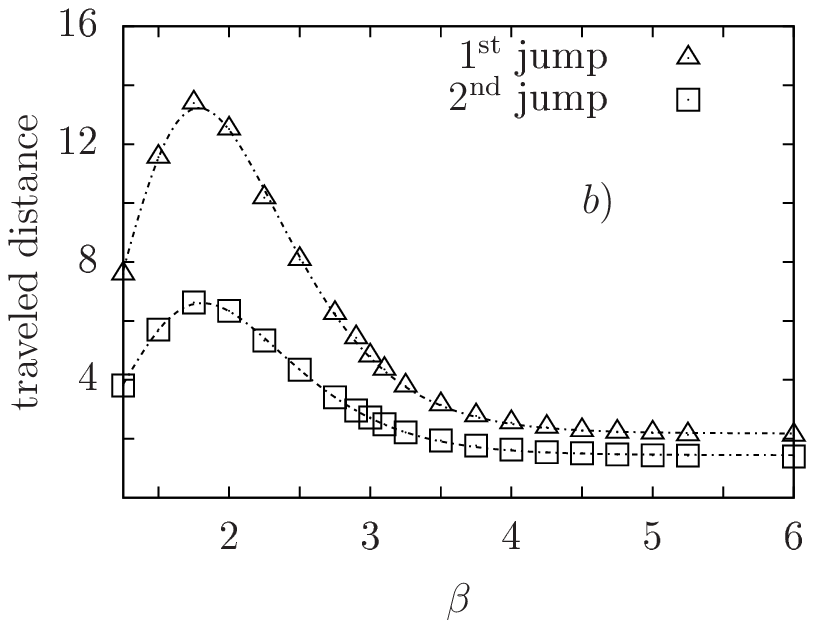}
\caption{Properties of the first and second jumps in  the $o2p$ model.  $a)$ Fraction of captured resources. $b)$ Total distance travelled. The lines are visual guides to the eye.}
\label{fig:sep2opaso}
\end{figure}

Figure  \ref{fig:sep2opaso}{\bf $a)$} displays the average fraction of captured resources $\langle K_T\rangle$ as given by $o2p$,  whereas Figure \ref{fig:sep2opaso}{\bf $b)$} shows the average total traveled distance $\langle l_T\rangle$.  Again, we have separated the contributions of the first and  second jumps to these quantities. The two types of jumps are clearly different. Interestingly, the second jumps are not only the ones that visit the more important targets, but they are also the shorter ones. The picture that emerges from these results is that the first jump of the two-step optimization serves as an ``approach'' towards some important targets, which would not be visited directly in the one-step optimization. 


\section{Discussion and conclusions} \label{disscus}

In this paper we have presented a computer model which mimics the mobility of an intelligent forager that has a perfect  knowledge of its environment. The model is ecologically inspired by the way many animals and humans interact with complex landscapes of resources, and it incorporates two important features: optimization (least effort rule) and multiple-step planning.  

We  found  that the average foraging efficiency obtained from a two-step optimization is larger compared to that the one-step optimization. Nevertheless, the calculation time needed in the former is orders of magnitude larger. In the two-step scenario, power-law step-length distributions emerge in certain heterogeneous environments which are characterized by a resource exponent $\beta$ near $3$, similarly to the case of the one-step process. We also found that the incorporation of more than one step in the planed movements decreases the fluctuations of the lengths of the resulting steps, or, in other words, diminishes the uncertainty faced by the forager. In the limiting case of $N$-step planning, the TSP is recovered.

The increased computing time for the two-step model does not bring a significant improvement in the forager efficiency, compared with the simpler single-step optimization. This result resonates with recent empirical findings on primate systems, where no evidence for multi-step planning was found \cite{janson2014death}. However, this issue is not completely settled, as other studies suggest that multiple-step planning is sometimes used by primates \cite{valero2007spider}. Some experiments even provide evidence that bumblebees are able to solve the TSP by learning from experience, but in simple configurations containing a small number of targets \cite{lihoreau2010travel}. We speculate that when the environment is very heterogeneous and offers many possibilities to the forager, one-step strategies may be sufficient and could have been selected through evolution.

We have found a clear asymmetry between the properties of the first and the second jump in the two-step optimization. The  first jump is usually longer, while most of the resources are captured in the second jump. Therefore, the function of the first step is to approach the forager closer to a large resource patch, that would otherwise not necessarily be chosen in a single-step process. This combination produces an increase in the foraging efficiency.

The foregoing results pose several challenging questions regarding the amount of multiple-step planning that humans are willing to perform in order to solve foraging, shopping, visiting or delivery problems. We suggest that it is unlikely that high order step optimization will be performed. This result could be generic and transcend the context of the present study. We speculate that single step processes may explain, for instance, why retail stores of the same kind tend to group in the same area, as close to each other as possible \cite{Jensen}.

\section{acknowledgments}  The   authors  appreciate  financial   support  from
DGPA-PAPIIT  Grant  IN105015 and PIIF2016-IFUNAM.    MM thanks  support
from the CONACyT Postgraduate Grants Program. OM  thanks financial support  from  a PASPA-UNAM Grant supporting a sabbatical leave at the Universidad Polit\'{e}cnica de Madrid, in Spain. OM is also grateful for the hospitality of the Mathematics Department at the ETSI-Aeronauticos at the UPM, Spain.


\end{document}